\documentclass{article}
\usepackage{amsmath}
\usepackage{appendix}
\usepackage[retainorgcmds]{IEEEtrantools}
\usepackage{graphicx}
\begin{document}
\title{Bogoliubov theory of interacting bosons: new insights from an old problem}
\author{Loris Ferrari \\ Department of Physics and Astronomy of the University (DIFA) \\Via Irnerio, 46,40127, Bologna, Italy}
\maketitle

\begin{abstract}
In a gas of $N$ interacting bosons, the Hamiltonian $H_c$, obtained by dropping all the interaction terms between free bosons with moment $\hbar\mathbf{k}\ne\mathbf{0}$, is diagonalized exactly. The resulting eigenstates $|\:S,\:\mathbf{k},\:\eta\:\rangle$ depend on two discrete indices $S,\:\eta=0,\:1,\:\dots$, where $\eta$ numerates the \emph{quasiphonons} carrying a moment $\hbar\mathbf{k}$, responsible for transport or dissipation processes. $S$, in turn, numerates a ladder of \textquoteleft vacua\textquoteright$\:|\:S,\:\mathbf{k},\:0\:\rangle$, with increasing equispaced energies, formed by boson pairs with opposite moment. Passing from one vacuum to another ($S\rightarrow S\pm1$), results from creation/annihilation of new momentless collective excitations, that we call \emph{vacuons}. Exact quasiphonons originate from one of the vacua by \textquoteleft creating\textquoteright$\:$an asymmetry in the number of opposite moment bosons. The well known Bogoliubov collective excitations (CEs) are shown to coincide with the exact eigenstates $|\:0,\:\mathbf{k},\:\eta\:\rangle$, i.e. with the quasiphonons created from the lowest-level vacuum ($S=0$). All this is discussed, in view of existing or future experimental observations of the vacuons (PBs), a sort of bosonic Cooper pairs, which are the main factor of novelty beyond Bogoliubov theory.\newline
\end{abstract}

e-mail: loris.ferrari@unibo.it
telephone: ++39-051-2095109\newline
\\

In his approach to the weakly interacting bosons gas \cite{Bogo1,Bogo2}, Bogoliubov's first step was eliminating from the $N$-bosons Hamiltonian:

\begin{equation}
H_{bos}=\sum_{\mathbf{k}}\overbrace{\left(\hbar^2k^2/2M\right)}^{\mathcal{T}(k)}b^\dagger_{\mathbf{k}}b_{\mathbf{k}}+\frac{1}{2}\sum_{\mathbf{k}_1,\mathbf{k}_2,\mathbf{q}}\widehat{u}(q)\:b^\dagger_{\mathbf{k}_2-\mathbf{q}}b^\dagger_{\mathbf{k}_1+\mathbf{q}}b_{\mathbf{k}_1}b_{\mathbf{k}_2}\:\label{Hbos}
\end{equation}
\\
all the interaction terms that couple bosons in the excited states. This yields the truncated canonic Hamiltonian:

\begin{align}
H_c&=\overbrace{\frac{\widehat{u}(0)N^2}{2}}^{E_{in}}+\sum_{\mathbf{k}\ne0}\overbrace{\left[\mathcal{T}(k)+\widetilde{N}_{in}\:\widehat{u}(k)\right]}^{\widetilde{\epsilon}_1(k)}b^\dagger_{\mathbf{k}}b_{\mathbf{k}}+\nonumber\\
\nonumber\\
&+\frac{1}{2}\sum_{\mathbf{k}\ne0}\widehat{u}(k)\Big[b^\dagger_{\mathbf{k}}b^\dagger_{-\mathbf{k}}(\:b_{\mathbf{0}}\:)^2+b_{\mathbf{k}}b_{-\mathbf{k}}(\:b^\dagger_{\mathbf{0}}\:)^2\Big]\:,\label{Hc}
\end{align}
\\ 
in the thermodynamic limiy (TL). The operators $b^\dagger_{\mathbf{k}}$ and $b_{\mathbf{k}}$ create/destroy a spinless boson in the free-particle state $\langle\:\mathbf{r}\:|\:\mathbf{k}\:\rangle= e^{i\mathbf{k}\:\mathbf{r}}/\sqrt{V}$
and  

\begin{equation*}
\widehat{u}(q)=\frac{1}{V}\int\mathrm{d}\mathbf{r}e^{-i\mathbf{q}\:\mathbf{r}}\:u(r)\:,
\end{equation*}
\\
is the Fourier transform of the \emph{repulsive} interaction energy $u(r)$ ($>0)$. The number operator $\widetilde{N}_{in}=b^\dagger_{\mathbf{0}}b_{\mathbf{0}}$ refers to the bosons in the free particle ground state. Accordingly, $\widetilde{N}_{out}=N-\widetilde{N}_{in}$ is the number operator for bosons in the excited states, $N$ being the conserved total number of bosons. Overtilded symbols indicate operators, to avoid confusion with their (non overtilded) eigenvalues.

Bogoliubov's next step is reducing $H_c$ to a bi-linear form, which is realized in the TL, by assuming $|\:N_{in}\pm2,\:N_{out}\:\rangle\approx|\:N_{in},\:N_{out}\:\rangle$ for the bosonic Fock states \cite{MS}. This yields the approximated Hamiltonian:

\begin{subequations}
\begin{align}
H_{BCA}&=E_{in}+\sum_{\mathbf{k}\ne0}\overbrace{\left[\mathcal{T}(k)+N\:\widehat{u}(k)\right]}^{\epsilon_1(k)}\beta^\dagger_{\mathbf{k}}\beta_{\mathbf{k}}+\nonumber\\
&+\frac{N}{2}\sum_{\mathbf{k}\ne0}\widehat{u}(k)\Big[\beta^\dagger_{\mathbf{k}}\beta^\dagger_{-\mathbf{k}}+\beta_{\mathbf{k}}\beta_{-\mathbf{k}}\Big]\:,\label{HBCA}
\end{align}
\\
where new creation/annihilation operators are introduced:

\begin{equation}
\label{beta,beta*}
\beta_{\mathbf{k}}=b_{\mathbf{0}}^\dagger\left(\widetilde{N}_{in}+1\right)^{-1/2}b_\mathbf{k}\quad,\quad\beta_{\mathbf{k}}^\dagger=b_{\mathbf{k}}^\dagger\left(\widetilde{N}_{in}+1\right)^{-1/2}b_\mathbf{0}\:,
\end{equation}
\end{subequations}
\\
ensuring the conservation of the number $N$ of real bosons. Note that $\beta_\mathbf{k}$ and $\beta_\mathbf{k}^\dagger$ satisfy the bosonic commutation rules \emph{exactly} \cite{AB, AB2}. This is what we call the Bogoliubov Canonic Approximation (BCA). Since $H_{BCA}$ is a bi-linear form in bosonic creation/annihilation operators, it is possible to eliminate the interactions by a suitable Bogoliubov transformation:

\begin{equation}
\label{B}
B^\dagger_{\mathbf{k}}=w^*_+\beta_{\mathbf{k}}^\dagger-w^*_-\beta_{\mathbf{-k}}\quad;\quad B_{\mathbf{k}}=w_+\beta_{\mathbf{k}}-w_-\beta_{\mathbf{-k}}^\dagger\:,
\end{equation}
\\
which diagonalizes $H_{BCA}$ as:

\begin{subequations}
\label{HBCAdiag,Epsilon}
\begin{equation}
\label{HBCAdiag}
H_{BCA}=\sum_{\mathbf{k}}\left[\left(B^\dagger_{\mathbf{k}}B_{\mathbf{k}}+1/2\right)\epsilon(k)-\frac{\epsilon_1(k)}{2}\right]\:,
\end{equation}
\\
with
\begin{equation}
\label{epsilon}
\epsilon(k)=\sqrt{\epsilon_1^2(k)-N^2\widehat{u}^2(k)}\:,
\end{equation}
\end{subequations}
\\
$\epsilon_1(k)$ being defined in eq.n~\eqref{HBCA}. Equations \eqref{HBCAdiag,Epsilon} correspond to a \emph{free} gas of new \textquoteleft particles\textquoteright, which is costumary to call \emph{collective excitations} (CEs), or \textquoteleft quasiparticles\textquoteright, created/annihilated by $B^\dagger_{\mathbf{k}}$, $B_{\mathbf{k}}$. In particular, the CEs resulting from Bogoliubov's approach are called \textquoteleft quasiphonons\textquoteright$\:$(QPs), due to their close similarity with the elastic modes of a solid. Actually, QPs behave as massless bosons, carrying a finite moment $\hbar\mathbf{k}$ and an energy $\epsilon(k)$ (eq.n~\eqref{epsilon}). The eigenvalues of $H_{BCA}$ are, obviously:

\begin{equation}
\label{EBCAS}
 E_{BCA}(S,\:k)=\left[\epsilon(k)\left(S+\frac{1}{2}\right)-\frac{\epsilon_1(k)}{2}\right]\quad (S=0,\:1,\:\dots)\:,
\end{equation}
\\
resulting from the creation of $S$ QPs.\\

In a series of papers \cite{MS, MS2, EC, MSArX}, I performed the \emph{exact} diagonalization of $H_c$, which showed some non trivial differences from the currently accepted theory described above. The first one is the existence of a new kind of CEs, which I called \textquoteleft pseudobosons\textquoteright$\:$in ref.~\cite{MS2}, but could be better denoted as \textquoteleft vacuons\textquoteright, as we shall see in what follows. They have been found in  ref.~\cite{ MS}, since now on referred to as (I), by diagonalizing the Hamiltonian \eqref{Hc} exactly, \emph{in the special subspace spanned by the states $|\:j,\:\mathbf{k}\:\rangle_0$ with the same number $j$ of bosons in $|\:\mathbf{k}\:\rangle$ and $|\:-\mathbf{k}\:\rangle$ and $N-2j$ bosons in $|\:\mathbf{0}\:\rangle$}:

\begin{equation}
\label{| j >0}
|\:j,\:\mathbf{k}\:\rangle_0=\frac{(b_\mathbf{0}^\dagger)^{N-2j}}{\sqrt{(N-2j)!}}\frac{(b_\mathbf{k}^\dagger)^j(b_\mathbf{-k}^\dagger)^j}{j!}|\emptyset\:\rangle\:
\end{equation}
\\
($|\emptyset\:\rangle$ being the real bosons' vacuum). The resulting eigenvalues $E_S(k,0)$ (see, in particular, \cite{EC}) turn out to be \emph{twice as large as} the BCA energies eq.n~\eqref{EBCAS}, reported in the current literature \cite{PS}:

\begin{equation}
\label{EpsilonS}
E_S(k,0)=2\underbrace{\left[\epsilon(k)\left(S+\frac{1}{2}\right)-\frac{\epsilon_1(k)}{2}\right]}_{E_{BCA}(S,\:k)}\quad (S=0,\:1,\:\dots)\:,
\end{equation}
\\
The exact eigenstates of $H_c$ calculated in (I) are, in turn, quite different from the BCA eigenstates: the latter have total moment $S\hbar\mathbf{k}$, corresponding to a number $S$ of QPs, while the formers are superpositions of free bosons states, with opposite moments, carrying a \emph{momentless} energy $E_S(k,0)$. After a private communication, I was informed that the vacuons had been actually discovered years ago, by J. Dziarmaga and K. Sacha (DS) \cite{Pol, Pol2}, by diagonalizing $H_c$ in a Fock subspace, formed by \emph{pairs} of bosons occupying the \emph{same} single-particle state. The not so relevant difference is that DS generalize the single-particle states to include possible external potentials (harmonic well, double or triple well). More importantly, DS do not solve the \emph{asymmetric} occupation case, as it is done in the present case. So, they do not realize the \textquoteleft vacuonic\textquoteright$\:$nature of the exact solutions they did find. Indeed, the physical meaning of the new CEs, and their relationship with the \emph{exact} QPs, follow from the complete diagonalization of $H_c$, including the subspaces containing a different number of bosons with opposite moment:

\begin{equation}
\label{| j >eta}
|\:j,\:\mathbf{k}\:\rangle_\eta=\frac{(b_\mathbf{0}^\dagger)^{N-2j-\eta}}{\sqrt{(N-2j-\eta)!}}\frac{(b_\mathbf{k}^\dagger)^{j+\eta}(b_\mathbf{-k}^\dagger)^j}{\sqrt{j!(j+\eta)!}}|\emptyset\:\rangle\:,
\end{equation}
\\
with $j+\eta$ bosons in $|\:\mathbf{k}\:\rangle$, $j$ bosons in $|\:-\mathbf{k}\:\rangle$ and $N-2j-\eta$ bosons in $|\:\mathbf{0}\:\rangle$, so that the total moment is, manifestly, $\eta\hbar\mathbf{k}$. I performed such diagonalization in refs~\cite{MS2, MSArX}, by using the method outlined below.\\ 

Hamiltonian $H_c$ can be written as a sum of independent one-moment Hamiltonians

\begin{subequations}
\begin{equation}
\label{Hc2}
H_{c}=E_{in}+\sum_{\mathbf{k}\ne0}h_c(\mathbf{k})\:,
\end{equation}
\\
where:

\begin{align}
h_c(\mathbf{k})&=\frac{1}{2}\widetilde{\epsilon}_1(k)[b^\dagger_{\mathbf{k}}b_{\mathbf{k}}+b^\dagger_{-\mathbf{k}}b_{-\mathbf{k}}]+\nonumber\\
\nonumber\\
&+\frac{1}{2}\widehat{u}(k)\Big[b^\dagger_{\mathbf{k}}b^\dagger_{-\mathbf{k}}(\:b_{\mathbf{0}}\:)^2+b_{\mathbf{k}}b_{-\mathbf{k}}(\:b^\dagger_{\mathbf{0}}\:)^2\Big]\:.\label{hc}
\end{align}
\end{subequations}
\\
Hence the whole problem can be reduced to the study of the exact eigenstates of $h_c(\mathbf{k})$, by solving the eigenvalue equation

\begin{equation}
\label{| S >eta}
h_c|S,\:\mathbf{k},\:\eta\:\rangle=\mathcal{E}_S(k,\eta)|S,\:\mathbf{k},\:\eta\:\rangle\:,
\end{equation}
\\
with exact QPs  expressed as linear combinations of the states eq.n~\eqref{| j >eta}. Thanks to a suitable transformation \cite{MSArX}, eq.n~\eqref{| S >eta} can be reduced to the same problem already solved \emph{analytically} in (I). This makes it possible to write the eigentates of $h_c$ (eq.n~\eqref{hc}) as :

\begin{equation}
\label{|S,eta>2}
|S,\:\mathbf{k},\:\eta\:\rangle=\sum_{j=0}^\infty \overbrace{x^j(k)\sqrt{\binom{j+\eta}{j}}\sum_{m=0}^SC_{S,\eta}(m,k)j^m}^{\phi_{S,\eta}(j,k)}|j,\:\mathbf{k}\:\rangle_\eta\:,
\end{equation}
\\
with boundary conditions $\mathrm{lim}_{j\rightarrow\infty}\phi_{S,\eta}(j,k)=0$ (necessary for normalizability) and $\phi_{S,\eta}(-1,k)=0$ (exclusion of negative populations). It should be noticed that $\left|\phi_{S,\eta}(j,k)\right|^2\propto j^{2S+\eta}x^{2j}(k)$ for $j>>1$, i.e. the pre-exponential factor in the probability amplitude on the Fock states $|j,\:\mathbf{k}\:\rangle_\eta$ (eq.n~\eqref{| j >eta}) tends to a polinomial of degree $2S+\eta$ in $j>>1$. The quantity $x(k)$, the coefficients $C_{S,\eta}(m,k)$ and the eigenvalue $\mathcal{E}_S(k,\eta)$ are determined by the following system of $S+2$ equations (see ref.~\cite{MSArX}):

\begin{align}
&\left(\underline{\epsilon}_1\eta-2\underline{\mathcal{E}}\right)C(n)+ 2\underline{\epsilon}_1C(n-1)+\nonumber\\
\nonumber\\
&+x\left[(1+\eta)\sum_{m=n}^SC(m)\binom{m}{n}+\sum_{m=n-1}^SC(m)\binom{m}{n-1}\right]-\nonumber\\
\nonumber\\
&+\frac{1}{x}\sum_{m=n-1}^SC(m)\binom{m}{n-1}(-1)^{n-m+1}=0\:\:(n=0,\:1,\:\dots,\:S+1)\:,\label{eq.C}
\end{align}
\\
where $C(S+1)=C(-1)=0$ by definition and the dependence on $\mathbf{k}$, $S$, $\eta$ has been provisionally dropped. The energies are expressed in units of $N\widehat{u}(k)$: 

\begin{equation*}
\label{underline}
\underline{\epsilon}_1=\frac{\epsilon_1(k)}{N\widehat{u}(k)},\:\underline{\mathcal{E}}=\frac{\mathcal{E}_S(k,\eta)}{N\widehat{u}(k)},\:\underline{\epsilon}=\frac{\epsilon(k)}{N\widehat{u}(k)}\:.
\end{equation*}
\\
The algebraic structure of the eigenvalue problem~\eqref{eq.C} is fairly peculiar: the two highest order equations ($n=S+1$ and $n=S$) determine $x(k)$ (i.e. the exponential decay in $j$) and the eigenvalue $\underline{\mathcal{E}}$, \emph{independently} from $C(S)$ and $C(S-1)$:

\begin{subequations}
\begin{equation}
\label{x=}
x(k)=\underline{\epsilon}(k)-\underline{\epsilon}_1(k)=\frac{\epsilon(k)-\epsilon_1(k)}{N\widehat{u}(k)}\:
\end{equation}
\\
\begin{align}
\mathcal{E}_S(k,\:\eta)&=\frac{\epsilon(k)}{2}\left(\eta+2S\right)-\nonumber\\
\nonumber\\
&-\frac{\epsilon_1(k)-\epsilon(k)}{2}\quad(S,\:\eta=0,\:1,\:\dots)\label{E(S,eta)}\:,
\end{align}
\end{subequations}
\\
with all entries restored. Notice that $x(k)$ is negative and smaller than 1 in modulus, which ensure normalizability. The next $S$ equations determine the $S$ unknowns $C(1),\:C(2),\dots,\:C(S)$ in terms of, say, $C(0)$, that will be determined by normalization. Figure 1 shows the resulting probability amplitudes $\left|\phi_{S,\eta}(j,k)\right|^2$, for $S=0,\:1,\:2$ and $\eta=0$ (1(a)), or  $\eta=1$ (1(b)). Since $h_c(\mathbf{k})=h_c(-\mathbf{k})$ (eq.n~\eqref{hc}), $\mathcal{E}_S(k,\:\eta)$ must be counted twice in the sum eq.n~\eqref{Hc2}. Hence the energy eigenvalues of $H_c-E_{in}$ are:

\begin{subequations}
\label{ES(k,eta)}
\begin{align}
&E_S(k,\:\eta)=2\mathcal{E}_S(k,\:\eta)=\epsilon(k)\left(\eta+2S\right)-\left[\epsilon_1(k)-\epsilon(k)\right]\label{E(S,eta)0}\\
\nonumber\\
&=\epsilon(k)\left(\eta+\frac{1}{2}\right)+2\epsilon(k)\left(S+\frac{1}{2}\right)-\underbrace{\left[\epsilon_1(k)+\frac{\epsilon(k)}{2}\right]}_{\mathcal{E}_0(k)}\label{E(S,eta)1}\:.
\end{align}
\end{subequations}
\\

It is useful to recall that the coefficients $w_\pm(\mathbf{k})$ in the Bogoliubov transformation eq.n~\eqref{B}, can be expressed in terms of $x(\mathbf{k})$ (eq.n~\eqref{x=}) as follows:

\begin{equation}
\label{w+-}
w_+=\frac{1}{\sqrt{1-x^2}} \quad;\quad w_-=\frac{x}{\sqrt{1-x^2}}\:.
\end{equation}
\\
The limit of large $k>>\xi^{-1}\equiv\sqrt{2M\widehat{u}(0)N}/\hbar$, that marks the passage from collective to single-particle dynamics, yields $\epsilon_1(k)\rightarrow\mathcal{T}(k)$ and $\epsilon(k)-\epsilon_1(k)\rightarrow0$. Hence, from eq.n~\eqref{ES(k,eta)}, one has:

\begin{equation}
\label{k>>}
E_S(k,\:\eta)\rightarrow(2S+\eta)\mathcal{T}(k)\quad(k>>\xi^{-1})\:.
\end{equation}
\\
Since the CEs become non interacting \emph{real} bosons when their kinetic energy $\mathcal{T}(k)=\hbar^2k^2/(2M)$ largely exceeds the interaction energy, the number $2S+\eta$ corresponds to the total number of real bosons excited in the limit $k>>\xi^{-1}$. This suggests the physical interpretation of the exact results obtained so far: $S$ numerates a sort of bosonic Cooper pairs, formed by \emph{two} bosons in $|\:\pm\mathbf{k}\:\rangle$, with opposite moments and identical kinetic energy. In contrast, $\eta$ numerates the additional bosons in $|\:\mathbf{k}\:\rangle$, carrying the non vanishing moment $\hbar\mathbf{k}$. Hence the states $|S,\:\mathbf{k},\:0\:\rangle$ do not carry any quasiparticle: each of them is nothing but a possible \textquoteleft vacuum\textquoteleft$\:$of QPs. This explains why the transition $S\rightarrow S\pm1$ is conveniently defined as the creation/annihilation of a quantum of vacuum (\emph{vacuon}), on which the QPs can be created in turn.\newline
\\

What precedes marks a relevant difference with respect to Bogoliubov's theory, where the QPs are obtained by repeated application of the creation operator $B^\dagger_{\mathbf{k}}$ (eq.n~\eqref{B}) to a \emph{single} vacuum state. Therefore, the BCA eigenvalues of the number operator $B^\dagger_{\mathbf{k}}B_{\mathbf{k}}$ are assumed to be \emph{non degenerate}, in as much as they represent the number of a single species of quasiparticles. As a consequence, the BCA spectrum corresponds to a \emph{single, non degenerate} oscillator, with frequency $\omega_Q(k)=\epsilon(k)/\hbar$ (recall $\mathcal{E}_S(k,0)$, in eq.n~\eqref{EpsilonS}). In contrast, recalling eq.ns\eqref{B} and \eqref{w+-}, a lengthy but straightforward calculation leads to the following exact result:

\begin{equation*}
B^\dagger_{\mathbf{k}}B_{\mathbf{k}}|\:S,\mathbf{k},\:\eta\:\rangle=(\eta+S)|\:S,\mathbf{k},\:\eta\:\rangle
\end{equation*}
\\
showing that: \newline
\\
(A) The $M$-th eigenvalue of $B^\dagger_{\mathbf{k}}B_{\mathbf{k}}$ is $M+1$-times \emph{degenerate}, since $M=S+\eta$ includes both the QPs and the vacuons number. The exact energy spectrum is degenerate too, resulting from the sum of \emph{two} oscillators (eq.n~\eqref{E(S,eta)1}), one labeled by the non negative integer $\eta$, with the same frequency $\omega_Q(k)$ as BCA, the other, labeled by the non negative integer $S$, with doubled frequency $\omega_P(k)=2\epsilon(k)/\hbar$.\newline

It is worthwhile noticing that $B^\dagger_{\mathbf{k}}B_{\mathbf{k}}$ still preserves its meaning of number operator, counting the \emph{total} number of CEs (QPs + vacuons). However, despite $S$ and $\eta$ can be varied independently, $B^\dagger_{\mathbf{k}}B_{\mathbf{k}}$ cannot be split into two distinct operators, counting QPs and vacuons separately. This is because, in general, $B^\dagger_{\mathbf{k}}$ and $B_{\mathbf{k}}$ are no longer creation/annihilation operators, though their product behaves like a number operator. The only exception is the case of zero vacuons $S=0$: by solving the system of equations for the $C_{S,\eta}(m,\:k)$ explicitly \cite{MSArX}, in fact, it is possible to show that

\begin{equation}
\label{B+|eta>}
B^\dagger_{\mathbf{k}}|0,\:\mathbf{k},\:\eta\:\rangle=\sqrt{\eta+1}|0,\:\mathbf{k},\:\eta+1\:\rangle\:.
\end{equation}
\\
Since in (I) it was shown that the lowest-order vacuum $|0,\:\mathbf{k},\:0\:\rangle$ coincides with the vacuum of Bogoliubov QPs, equation~\eqref{B+|eta>} shows that Bogoliubov QPs coincide with the exact QPs, in the absence of vacuons. In this special case SBA lead to the exact CEs $|\:0,\mathbf{k},\:\eta\:\rangle$ (eq.n~\eqref{|S,eta>2}), corresponding to $\eta$ QPs, each carrying a moment $\hbar\mathbf{k}$, with energy $2\mathcal{E}_0(k,1)=\epsilon(k)$ and velocity $\mathbf{v}_Q(k)=\nabla_{\mathbf{k}}\omega_Q(k)$ (the \textquoteleft sound velocity\textquoteright). In the limit of small $k<<\xi^{-1}$, the sound velocity attains, in modulus, the constant value:

\begin{equation}
\label{soundvel}
v_Q=\sqrt{\frac{N\widehat{u}(0)}{M}}\:.
\end{equation}
\\
However:\newline 
\\
(B) The possible QP-vacua resulting from the exact diagonalization are all the eigenstates $|S,\:\mathbf{k},\:0\:\rangle$ ($S=0,\:1,\:\dots$). They originate from the activation of $S$ momentless energy quanta $\hbar\omega_P(\mathbf{k})$, that we call vacuons, and form an \emph{immobile} \textquoteleft sea\textquoteright$\:$of opposite moment pairs (Fig. 1(a)). The exact eigenstates $|S,\:\mathbf{k},\:\eta\:\rangle$ correspond to $\eta$ QPs, each with finite moment $\hbar\mathbf{k}$,  created from the $S$-th vacuum, as unpaired bosons producing an asymmetry in the opposite moment populations. \newline 
\\
Hence:\newline
\\  
(C) There are infinite different QPs, originated by the simoultaneous creation of $S\ne0$ vacuons  (Fig. 1(b)), whose velocity $\mathbf{v}_Q(k)=\nabla_{\mathbf{k}}\omega_Q(k)$ is the same as Bogoliubov's velocity, but whose energy is $\epsilon(k)(1+2S)$. Those quasiparticles are totally ignored by Bogoliubov's theory. In practice, SBA exactly reproduces just the part of the whole spectrum corresponding to $S=0$. \newline

\begin{figure}[htbp]
\begin{center}
\includegraphics[width=4in]{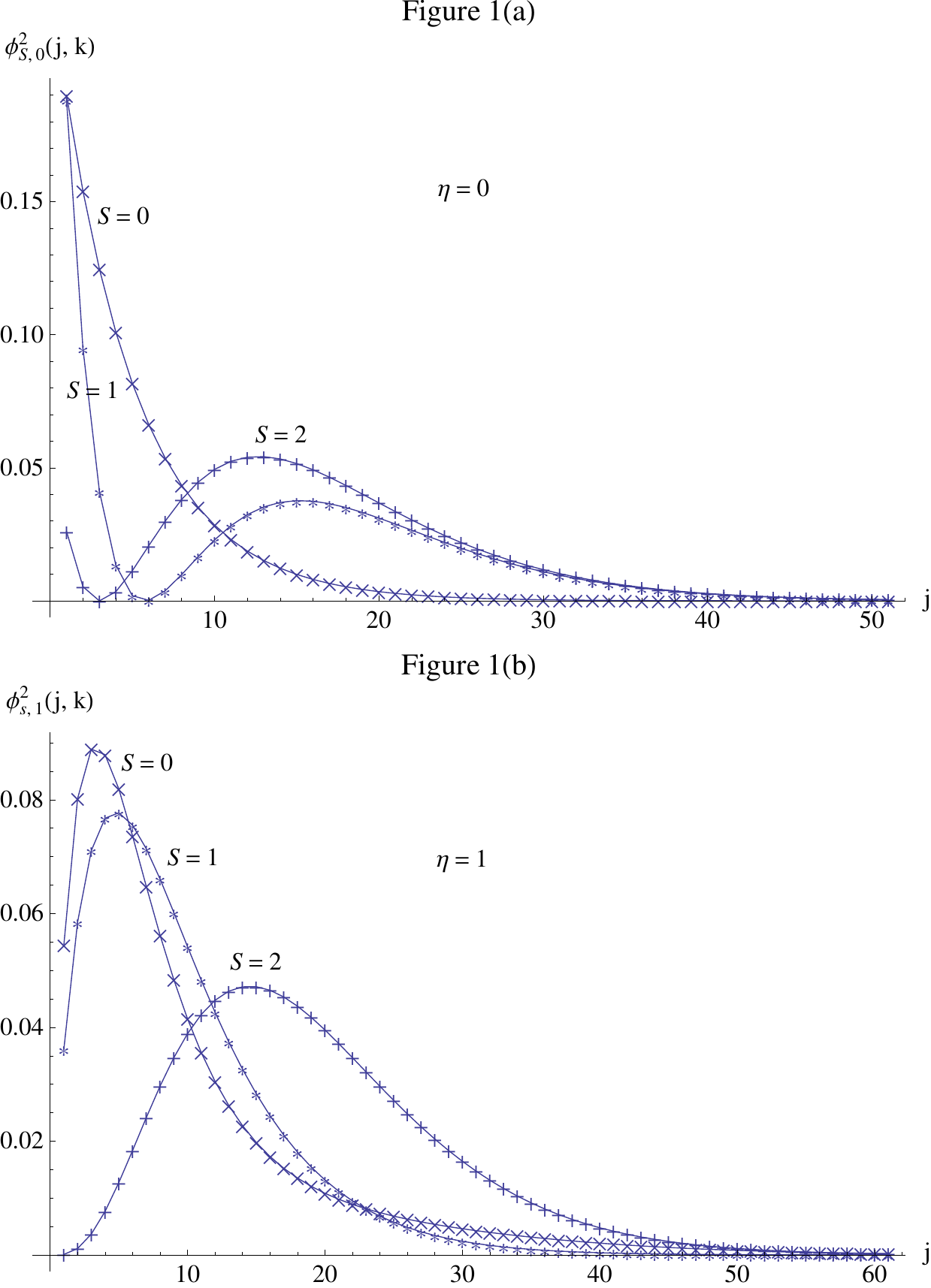}
\caption{\textbf{Probability amplitude of exact collective excitations ($x(\mathbf{k})=-0.9$).}\newline
\textbf{(a)}: CEs corresponding to QP-vacua ($\eta=0$), originated by creation of $S=0,\:1,\:2$ vacuons (bosonic Cooper pairs). \textbf{(b)}: CEs corresponding to 1 QP ($\eta=1$), created from $S=0,\:1,\:2$-level vacuum.\newline
$j$ is the number of opposite moment pairs in the subspace spanned by the states $|j,\:\mathbf{k}\:\rangle_0$ \textbf{(a)} (eq.n~\eqref{| j >0}) or $|j,\:\mathbf{k}\:\rangle_1$ \textbf{(b)} (eq.n~\eqref{| j >eta}). The collective excitation regime $k<<\xi^{-1}$ implies $1-|x(\mathbf{k})|<<1$.}
\label{default}
\end{center}
\end{figure}

Though the vacuons \emph{alone} cannot be responsible for the dissipation processes $\acute{a}$ \emph{la} Landau~\cite{L}, since they have zero moment, they can influence the dynamics of the dissipation processes, by opening new emission/adsorption channels for the QPs themselves. This can be easily seen by refreshing Landau's semiclassical picture of a particle (of mass $M_p$), moving with initial velocity $\mathbf{v}_i$ in  the superfluid at zero temperature, and causing the emission of a single QP, \emph{via} a scattering process. The well known conclusion, based on BCA, is that the particle cannot dissipate its kinetic energy (relative to the superfluid), unless $v_i>v_Q$ (eq.n~\eqref{soundvel}). Taking the superfluid in the ground state ($S=0$, $\eta=0$) initially, the exact expressions \eqref{ES(k,eta)} yield:

\begin{subequations}
\label{E-PCons}
\begin{align}
\frac{M_pv_i^2}{2}&=\frac{M_pv_f^2}{2}+\epsilon(k)(1+2S)\label{EnCons}\\
\nonumber\\
&M_p\mathbf{v}_i=M_p\mathbf{v}_f+\hbar\mathbf{k}\:,\label{MomCons}
\end{align}
\end{subequations}
\\
from energy/moment conservation, with $S$ vacuons and 1 QP in the final state of the superfluid. At small $k$, equations~\eqref{E-PCons} yield a succession of critical velocities

\begin{equation}
\label{vc(S)}
v_c(S)=v_Q(1+2S)\:\:(S=0,\:1,\:\dots)\:,
\end{equation}
\\
such that passing from $v_i< v_c(S)$ to $v_i> v_c(S)$ (or \emph{vice versa}) causes the switching on (off) of a new channel of QP emission, corresponding to the simultaneous creation (annihilation) of a vacuon. Note that this is not a standard multiparticle process, usually forbidden at first order. Actually, a straightforward calculation yields a non vanishing matrix element, in the TL, connecting the initial and final states of the process~\eqref{E-PCons}:

\begin{subequations}
\begin{align}
&\langle\:\mathbf{k}_i|\otimes\langle 0,\: \mathbf{q},\:0\:|U|S,\:\mathbf{q},\:\eta\:\rangle\otimes|\:\mathbf{k}_f\rangle=\nonumber\\
\nonumber\\
&=2W_0\:\delta_{\eta,1}\delta_{\Delta\mathbf{k},\mathbf{q}}\sqrt{N}\sum_{j=0}^\infty\phi_{0,0}(j,\mathbf{q})\phi_{S,1}(j,\mathbf{q})\sqrt{j+1}\:,\label{M-EL}
\end{align}
\\
where $\Delta\mathbf{k}=\mathbf{k}_i-\mathbf{k}_f$ and

\begin{equation}
\label{Usc}
U=\frac{W_0}{V}\sum_{\mathbf{k},\mathbf{k}'}\mathrm{e}^{i(\mathbf{k}-\mathbf{k}')\mathbf{r}}b_{\mathbf{k}'}^\dagger b_{\mathbf{k}}
\end{equation}
\end{subequations}
\\
represents a repulsive $\delta$-shaped interaction ($W_0>0$), between bosons in the superfluid and the scattering particle, with initial and final velocity $\mathbf{v}_{i,f}=\hbar\mathbf{k}_{i,f}/M_p$. Expression \eqref{Usc} adopts the 2nd quantization representation for the superfluid, and the co-ordinate ($\mathbf{r}$) representation for the scattering particle. Unlike the simultaneous creation/annihilation of  a \emph{single} QP and $S$ vacuons, it is easy to see that the creation/annhilation of 2 or more QPs is a multiparticle process, forbidden at first order. In fact:

\begin{equation}
\langle\:\mathbf{k}_i|\otimes\langle 0,\: \mathbf{q},\:0\:|U|S,\:\mathbf{q},\:\eta\:\rangle\otimes|\:\mathbf{k}_f\rangle=0\quad\text{if }\eta> 1\:.\label{M-EL=0}
\end{equation}
\\

As for the experimental verification of the new results obtained, double-photon Bragg spectroscopy is a current method to observe QPs \cite{S-K..., SOKD, UTR..., SBR...}, thanks to which it is possible to estimate the resonant values $\omega_{res}(k)$ of the difference between the two incident photons' frequencies, in the phonon regime $k<\xi^{-1}$, or in the free particle regime $k>\xi^{-1}$ (see, for example, Fig. 2 of ref.~\cite{S-K...}). The knowledge of $\omega_{res}(k)$ leads to the static structure factor $F_0(k)$ \cite{M}, from which, however, no difference can be revealed, between exact results and Bogoliubov theory. In fact, $F_0(k)$ involves just the ground state ($S=0,\:\eta=0$), which is the same in both cases, as shown in (I). However, a direct observation of the vacuons could stem from the measured $\omega_{res}(k)$s themselves. On a qualitative feet, in fact, the double oscillator spectrum eq.ns~\eqref{ES(k,eta)} should result in a succession of resonant frequencies $\omega_{res}=\omega(k)\left[\eta+2S\right]$, each corresponding to a hump in the transferred moment per particle. In particular, the activation of a single QP ($\eta=1$) with zero or one vacuon ($S=0,\:1$) is expected to produce \emph{two} humps at $\omega_{res}=\omega(k),\: 3\omega(k)$. The possible hump at $\omega_{res}=2\omega(k)$, corresponding to the creation of two QPs and zero vacuons, is expected to be a second-order effect, at most, in view of eq.n~\eqref{M-EL=0}. All the way, the data presently available (to the author's knowledge) do not cover a range of frequencies large enough to include the second (predicted) hump at $\omega_{res}=3\omega(k)$ (see Figure 2). Extending the measurements in ref.~\cite{S-K...} up to frequencies values of order $2\pi\times10^4Hz$ could provide a hint, in the interacting bosons regime ($k<\xi^{-1}$). Unfortunately, the data for the non interacting case ($k>\xi^{-1}$), as reported in Fig 2 of ref.~\cite{S-K...}, stop just below the value $3\omega(k)$, at which the second hump is expected to occur \cite{DS-K}.

\begin{figure}[htbp]
\begin{center}
\includegraphics[width=6in]{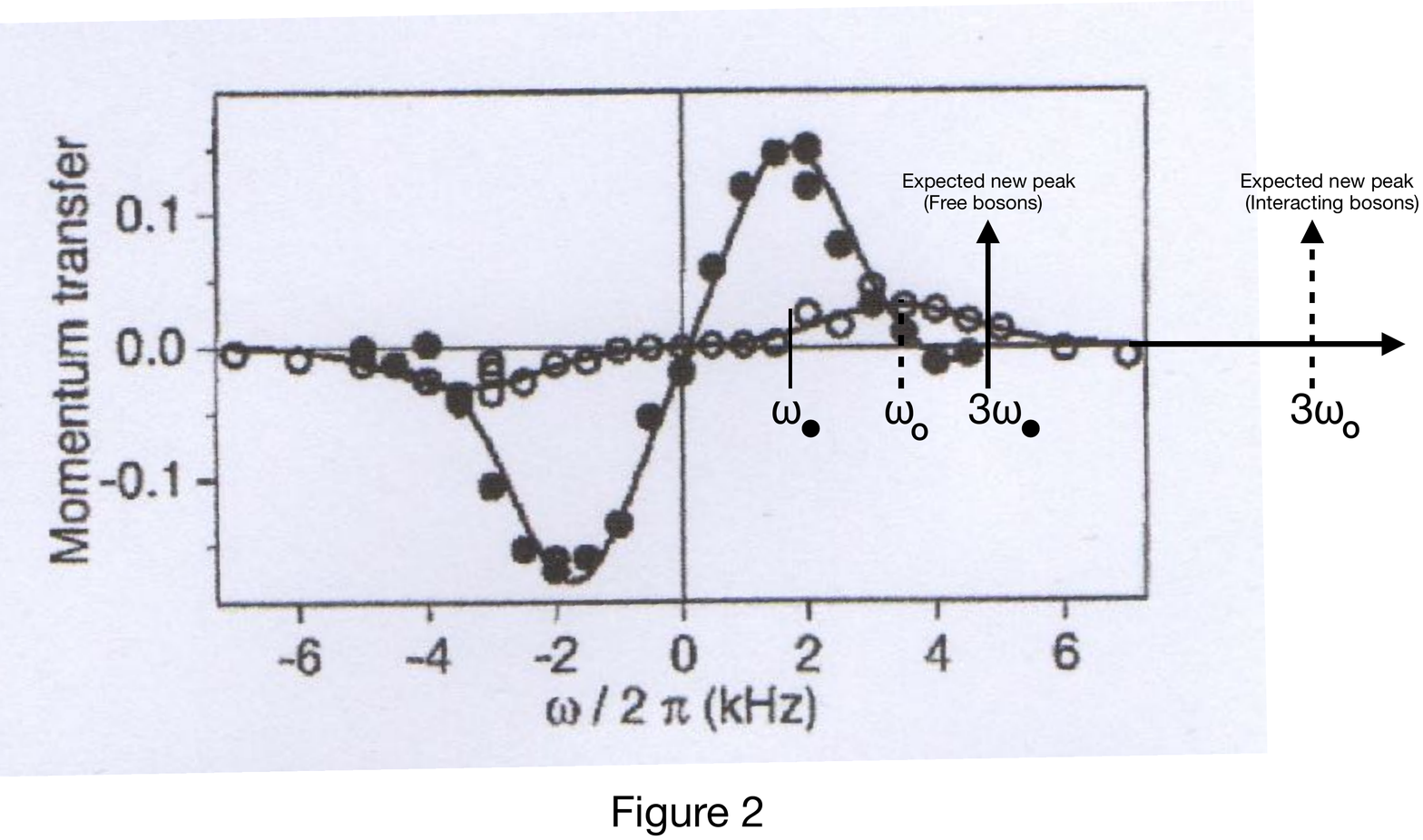}
\caption{\textbf{Measured momentum transfer from ref.~\cite{S-K...}.}\newline
Full and empty circles refer to free ($k>\xi^{-1}$) and interacting particle ($k<\xi^{-1}$) regime, respectively. The resonant frequencies $\omega_{\bullet}$ and $\omega_{\mathbf{o}}$ and the positions of the expected peaks at $3\omega_{\bullet}$ and $3\omega_{\mathbf{o}}$ are indicated accordingly. The antisymmetric part of negative frequencies is ignored.}
\label{default}
\end{center}
\end{figure}

In principle, the proliferation of critical velocities in eq.n~\eqref{vc(S)} could be observed by angle-resolved spectroscopy of bullet particles, impinging on a thin layered (single-scattering approximation) superfluid. Due to the switching on of a new channel of scattering, one would expect a hump in the flux of scattered particles at finite angle, whenever the incident beam's velocity crosses (from below) one of the critical values eq.n~\eqref{vc(S)}. The scattering particle velocities involved in experiments with $\mathrm{Na}$ \cite{S-K...} or $\mathrm{Rb}$ \cite{SOKD} atoms ($\approx10\:\mathrm{mm}/\mathrm{s}$) are unfeasible, in a real experiment, while those involved in exciton-polariton condensates ($\approx10^{7}\mathrm{cm/s}$ for electrons, $\approx10^{5}\mathrm{cm/s}$ for neutrons), as roughly estimated from data in ref.~\cite{UTR...}, look more accessible. 

Thermal depletion could provide an indirect evidence of the cohexistence of QPs and vacuons. From the exact energy~\eqref{E(S,eta)0}, one defines

\begin{equation}
S_T=\frac{1}{\mathrm{e}^{2\beta\epsilon(k)}-1}\quad,\quad \eta_T=\frac{1}{\mathrm{e}^{\beta\epsilon(k)}-1}
\end{equation}
\\
as the number of thermally activated vacuons and QPs. Then the exact thermal depletion follows from the calculated number of excited bosons in $|\:\pm\mathbf{k}\:\rangle$:
 
\begin{equation}
\label{n_k}
\langle\:n_{-\mathbf{k}}\:\rangle_T = \sum_{j=0}^\infty j\:\phi_{S_T,\eta_T}^2(j\:,k)\:\:,\:\:\langle\:n_{\mathbf{k}}\:\rangle_T= \sum_{j=0}^\infty j\:\phi_{S_T,\eta_T}^2(j\:,k)+\eta_T,
\end{equation}
\\
to be compared with measured values, as those reported, for instance, in ref.~\cite{SBR...}. However, in the experimental procedure and in the theoretical calculations of ref.~\cite{SBR...} there are a number of details that must be carefully accounted for, in view of a reliable comparison with eq.ns~\eqref{n_k}. This is a program for the future.\newline

In conclusion, the exact diagonalization of the truncated Hamiltonian $H_c$ (eq.n~\eqref{Hc}) reveals a hidden side of Bogoliubov collective excitations, i.e. the existence of an equispaced ladder of zero-point energies, each corresponding to an \emph{immobile} \textquoteleft sea\textquoteright$\:$ of boson pairs with opposite moment, on which QPs can be created/annihilated, as unpaired bosons producing an asymmetry in the opposite moment populations. Bogoliubov theory, based on SBA, accounts just for the QPs created from the lowest-level vacuum, but ignores all the higher-level CEs. Passing from one vacuum to another corresponds to the creation/annihilation of what we call \emph{vacuons}. Those CEs, reminiscent of bosonic Cooper pairs, are expected to produce observable effects, whose experimental verification represents a new challenging item for future investigations. 
\\

\end{document}